\documentclass{amsart}

\newcommand{\vet}[1]{{\ensuremath{\mbox{\boldmath $#1$}}}}
\usepackage{amssymb}
\usepackage{amsmath}
\usepackage[hypertexnames=false]{hyperref}  

\usepackage{amsthm}
\theoremstyle{plain}
\theoremstyle{definition}
\newtheorem{example}{Example}

\theoremstyle{remark}

\usepackage[numbers]{natbib}

\newcommand*{\QEDE}{\hfill\ensuremath{\blacksquare}}

\makeatletter
\newcommand{\addresseshere}{%
  \enddoc@text\let\enddoc@text\relax
}

\begin{document}

\title{Inequality Constrained Multilevel Models}

\author[B.S.\ Kato]{Bernet S.\ Kato}
\address[Bernet S.\ Kato]{
Population Health \& Occupational Disease \\
National Heart and Lung Institute \\
Imperial College \\
London \\
United Kingdom}

\author[C.F.W.\ Peeters]{Carel F.W.\ Peeters}
\address[Carel F.W.\ Peeters]{
Dept.\ of Epidemiology \& Biostatistics \\
Amsterdam Public Health research institute \\
VU University medical center Amsterdam \\
Amsterdam\\
The Netherlands}

\maketitle

\section{Multilevel Models} \label{ch11_sec1}
\subsection{Introduction}     \label{ch11_intro}
In many areas of research, datasets have a multilevel or
hierarchical structure. By hierarchy we
mean that units at a certain level are grouped or clustered into, or
nested within, higher-level units. The
``level'' signifies the position of a
unit or observation within the hierarchy. This implies that the data
are collected in groups or clusters. Examples of clusters are
families, schools, and firms. In each of these examples a cluster is
a collection of units on which observations can be made. In the case
of schools, we can have three levels in the hierarchy with pupils
(level 1) within classes (level 2) within schools (level 3). The key
thing that defines a variable as being a level is that its units can
be regarded as a random sample from a wider population of units. For
example, considering a multilevel data structure of pupils within classes within schools, the pupils are
a random sample from a wider population of pupils and the classrooms
are a random sample from a wider population of classrooms. Likewise
the schools are a random sample from a wider population of schools.
Data can then be collected at the pupil level (for example, a test
score), at the classroom level (for example, teacher experience in
years), and at the school level (for example, school's mean
socioeconomic status). Variables like gender and social class are
not levels. This is because they have
a small fixed number of categories. For example, gender has only two
categories, male and female. There is no wider population of gender
categories that male and female are a random sample from. Another
usual form of clustering arises when data are measured repeatedly on
the same unit, for instance a patient. In this case the measurements
from each patient would be at level 1 and the patients would be at
level 2.

In all cases the elements of a cluster share some common
characteristics. Therefore, the observations within a cluster tend
to be more alike than observations from different clusters, that is,
they are correlated. For instance, in the pupils within classrooms
example, pupils in the same classroom share some common
characteristics (e.g., they have the same teachers); thus the test
scores of pupils within a classroom will tend to be more alike than
test scores from different classrooms. Multilevel data therefore have two sources of
variation: In addition to the variation within clusters, the
heterogeneity between clusters introduces an additional source of
variation. Therefore, any analysis methods used should take the
\emph{within} cluster and \emph{between} cluster variation into
account. Because data can be clustered at more than a single level
(e.g., pupils within classrooms within schools), data clustered at a
single level (e.g., pupils within classrooms) are referred to as
two-level data and the statistical models for the analyses are
referred to as two-level models.

Multilevel or hierarchical data structures can occur in many areas of research, including
economics, psychology, sociology, agriculture, medicine, and public
health. Over the last 25 years, there has been increasing interest
in developing suitable techniques for the statistical analysis of
multilevel data, and this has resulted
in a broad class of models known under the generic name of
$multilevel$ models. Generally, multilevel models are useful for exploring how relationships vary across
higher-level units taking into account the \emph{within} and
\emph{between} cluster variations. Considering an example of
two-level data obtained on pupils
within schools, there are two possible ways to deal with the data:
either to focus separately on the pupils or on the schools. Focusing
on the pupils by \emph{pooling} together the data from all the
schools ignores differences between schools and thus suppresses
variation that can be important. Ignoring the clustering will
generally cause standard errors of regression coefficients to be
underestimated. On the other hand, focusing on schools by analyzing
the data of each school separately ignores a lot of information and
consequently renders low power for inferences. Multilevel modeling
offers a compromise between these two extremes and enables
researchers to obtain correct inferences.

\subsection{The Multilevel Model}   \label{ch11_mult}

In this chapter we will confine ourselves to two-level models for
continuous data, with one single outcome or response variable that
has been measured at the lowest level and explanatory variables (or
covariates) that have been measured at levels $1$ and $2$. For the
sake of consistency, level $1$ and level $2$ units will be referred
to as $\emph{individuals}$ and $\emph{groups}$, respectively. Stated
otherwise, \emph{individuals} will be nested within $\emph{groups}$.

To fix ideas, suppose we have $J$ groups and $N_j$ individuals in
each group such that the total number of individuals is $N$.
Furthermore, assume that one covariate $a$ has been measured at the
individual level and one covariate $w$ has been measured at the
group level and an outcome variable $y$ has been measured on each
individual. As an illustration suppose we have data on mathematics
grades from $N$ high school students from $J$ classes as well as
information on student socioeconomic background and teacher
experience in years. In this case, each of the classrooms would be a
$group$ and the students would be the $individuals$. Furthermore,
$y$ would be the student level outcome variable ``math grade,'' $a$
would be student ``socioeconomic status,'' and $w$ would be
``teacher experience'' in years. Our interest is in modeling the
outcome variable $y$ in terms of the individual level variable $a$
and the group level variable $w$ using a multilevel
model. At the individual level, for
individual $k$ (where $k=1,\ldots,N_j$ for group $j$) within group
$j$ ($j=1,\ldots,J$ groups in the sample) and $\sum_j N_j = N$, we
have the following model:
\begin{equation}
y_{kj} = \pi_{1j} + \pi_{2j}a_{kj} + \varepsilon_{kj}.
\label{leveloneModel}
\end{equation}
In (\ref{leveloneModel}), $\pi_{1j}$ is the intercept, $\pi_{2j}$ is
the regression coefficient for the covariate $a$, and $\varepsilon$
is the residual error term. The residual errors $\varepsilon_{kj}$
are assumed to have a normal distribution with mean $0$ and variance $\sigma^2$. Model
(\ref{leveloneModel}) implies that each group $j$ has its own
regression equation with an intercept $\pi_{1j}$ and a slope
$\pi_{2j}$. The next step in the
 modeling
 is to explain the variation of
the regression coefficients $\pi_{1j}$ and $\pi_{2j}$ by introducing
variables at group level:
\begin{eqnarray}
\pi_{1j}& = &\beta_{1} + \beta_{2}w_{j} + u_{1j},
\label{leveltwoModelInt} \\
\pi_{2j}& = &\beta_{3}+ \beta_{4}w_{j} + u_{2j},
 \label{leveltwoModelSlo}
\end{eqnarray}
where $u_{1j}$ and $u_{2j}$ are random residual error terms at group
level. Note that in (\ref{leveltwoModelInt}) and
(\ref{leveltwoModelSlo}), the regression coefficients ($\beta$'s) do
not vary across groups and that is why they have no subscript $j$ on
them. Since they apply to all groups, they are sometimes referred to
as $fixed$ effects. Furthermore, all between group variation left in
the $\pi$ coefficients after predicting them with the group variable
$w_j$ is assumed to be random residual variation (at group level)
which is captured by the terms $u_{1j}$ and $u_{2j}$.

Substituting (\ref{leveltwoModelInt}) and (\ref{leveltwoModelSlo})
into (\ref{leveloneModel}) renders the linear $two$-$level$
regression model:
\begin{equation}
y_{kj} = \beta_{1} + \beta_{2}w_{j} + \beta_{3}a_{kj} +
\beta_{4}a_{kj}w_{j} + u_{1j} + u_{2j}a_{kj} + \varepsilon_{kj}.
\label{twolevelmodel}
\end{equation}
The right-hand side of model (\ref{twolevelmodel}) has two parts to
it: a $fixed$ part $\beta_{1} + \beta_{2}w_{j} + \beta_{3}a_{kj} +
\beta_{4}a_{kj}w_{j}$, where the coefficients are fixed, and a
$random$ part $u_{1j} + u_{2j}a_{kj} + \varepsilon_{kj}$. Note that
in practice one can have several covariates measured at both
individual and group level. Therefore, model (\ref{twolevelmodel})
can be written in a slightly more general form using vector
notation:
\begin{equation}\label{vectwolevelmodel}
y_{kj} = \vet{x}_{kj}\vet{\beta}^T + \vet{z}_{kj}\vet{u}_{j}^T +
\varepsilon_{kj}, 
\end{equation}
where $\vet{x}_{kj}$ is a vector of predictors (including main
effects at levels 1 and 2 as well as interactions between level 1
and level 2 covariates) having coefficients $\vet{\beta}$.
Furthermore, $\vet{z}_{kj}$ is a vector of predictors having random
effects $\vet{u}_{j}$ at the group level and $\varepsilon_{kj}$ is
an error term. In the example above, $\vet{x}_{kj}$ =
($1,w_{j},a_{kj},a_{kj}w_{j}$), $\vet{z}_{kj}$= $(1,a_{kj})$,
$\vet{\beta}$ = ($\beta_{1},\beta_{2},\beta_{3},\beta_{4}$), and
 $\vet{u}_j =
(u_{1j},u_{2j})$. The vector of predictors $\vet{z}_{kj}$ will
usually be a subset of the fixed-effects predictors $\vet{x}_{kj}$,
although this is not a necessary requirement. The random terms
$\vet{u}_j = (u_{1j},u_{2j})$ and $\varepsilon_{kj}$ are assumed to
be mutually independent and normally distributed:
\begin{equation}
\vet{u}^T \sim \mathcal{N}(\vet{0},\vet{V}), ~~ \varepsilon_{kj}
\sim \mathcal{N}(0,\sigma^{2}), \label{randomterms}
\end{equation}
where $\vet{V}$ is the variance-covariance matrix of the random
effects and $\sigma^2$ is the residual variance. Thus, we can see
that multilevel models provide a natural way
to decompose complex patterns of variability associated with
hierarchical structure.

In a frequentist analysis, estimation of parameters in the linear
multilevel model is carried out by
maximizing the likelihood function. To this end, direct maximization
using the Newton-Raphson or Expectation-Maximization (EM) algorithm
can be performed. For discussions on the methods, techniques, and
issues involved in multilevel modeling in general, the interested
reader is referred to \cite{BRYR99, GEHIL07, GOLD95, HOX02, LONG93,
SWIL03, SNBOS99}. This chapter is intended to illustrate model
selection for inequality constrained two-level models. A Bayesian
approach will be used for parameter estimation and model
selection \cite{KHOI06}. Bayesian estimation
in multilevel models (without constraints on the model parameters)
has also been implemented in the statistical package MLwiN
\cite{BROWJ03}.

\section{Informative Inequality Constrained Hypotheses}
 \label{ch11_sec2}

Research scientists often have substantive theories in mind when
evaluating data with statistical models. Substantive theories often
involve inequality constraints among
the parameters to translate a theory into a model; that is, a
parameter or conjunction of parameters is expected to be larger or
smaller than another parameter or conjunction of parameters. Stated
otherwise and using $\beta$ as a generic representation of a
parameter, we have that $\beta_{i}>\beta_{j}$ or
$\beta_{i}<\beta_{j}$ for some two parameters $\beta_i$ and
$\beta_j$. Additionally, inequality constraints also play a pivotal role when competing theories are
presented as an expression of a multitude of initial plausible
explanations regarding a certain phenomenon on which data are
collected. Consider the following examples on two common multilevel
models: school effects models and
individual growth models. These
examples will be the thrust of Sections \ref{ch11_sec4} and
\ref{ch11_sec5}.

\begin{example}
An educational researcher
is interested in the effect of certain student and school level
variables on mathematical achievement
(\emph{mathach}), and has obtained a dataset on students within
schools. A students' ethnic background (\emph{min}), student
socioeconomic status (\emph{ses}), a schools' average student
socioeconomic status (\emph{mses}), and the dichotomy between
Catholic (\emph{cat}) and public (\emph{pub}) schools are
hypothesized to be defining variables for the explanation of math
achievement (cf. \cite{BEK93, BRYR99,
GAM92, GEAR94, SING98}). A possible formulation of the two-level
model in the form (\ref{vectwolevelmodel}) might be
\begin{eqnarray}\label{examp1}\nonumber
mathach_{kj} &=& \beta_{1}cat_{j}+\beta_{2}pub_{j}+\beta_{3}mses_{j}
+ \beta_{4}cat_{j}ses_{kj}\\\nonumber &&+
~\beta_{5}pub_{j}ses_{kj}+\beta_{6}mses_{j}ses_{kj} +
\beta_{7}min_{kj} \\\nonumber &&+ ~u_{1j} +
u_{2j}ses_{kj}+\varepsilon_{kj}.
\end{eqnarray}
The reason for assigning an indicator variable to both the Catholic
and public category of the constituent dichotomy is because this
will enable one to estimate the regression coefficients
corresponding to the covariates $cat$ and $pub$ and their
interactions with other covariates rather than estimating contrasts.

The researcher can think of different plausible models regarding the
direction and (relative) strength of the effects of the mentioned
variables on the response math achievement. Subsequently, the researcher expresses the idea that
students in Catholic schools have higher math
achievement than those in public
schools \{$\beta_{1}>\beta_{2}$\}. Certain sociological work found
that students belonging to a minority have lower math
achievement than students not belonging
to an ethnic minority \{$\beta_{7}<0$\}. Additionally, the
researcher has the expectation that math achievement is positively related to socioeconomic status and that
the effect of student socioeconomic status on mathematical
achievement is more pronounced in
public schools than in Catholic schools \{$\beta_{4}<\beta_{5}$\}.
These theories allow for several plausible models of differing
complexity and with differing theoretical implications. The question
of interest becomes: \emph{Which of the plausible models best fits
the data?}
\QEDE
\end{example}

\begin{example}
A researcher in child and
adolescent psychology has obtained observational data on substance
abuse collecting multiple waves of data on adolescents. This
researcher sets out to assess the effects of alcoholic intake among
peers (\emph{peer}) and the fact that the adolescent has alcoholic
({\emph{coa}}) or nonalcoholic (\emph{ncoa}) parents on the
development of adolescent alcohol use (\emph{alcuse}) (cf. \cite{CSC97, SWIL03}). The model can be
formulated as
\begin{eqnarray}\label{examp2}\nonumber
alcuse_{kj} &=&
\beta_{1}coa_{j}+\beta_{2}ncoa_{j}+\beta_{3}peer_{j}+
\beta_{4}coa_{j}t_{kj}+\beta_{5}ncoa_{j}t_{kj}\\\nonumber&&+
~\beta_{6}peer_{j}t_{kj} +u_{1j} +u_{2j}t_{kj} + \varepsilon_{kj},
\end{eqnarray}
where $t_{kj}$ is a time variable.

For these data, competing theories abound in the researchers' mind.
A first plausible theory for him or her could be that
adolescents with an alcoholic
parent are more prone to have a higher alcoholic intake at baseline
\{$\beta_{1}>\beta_{2}$\}, as well as over time
\{$\beta_{4}>\beta_{5}$\}. A second plausible theory amends the
first, with the additional expectation that for initial alcoholic
intake, the effect of an alcoholic parent will be more influential
than peer alcoholic intake \{$\beta_{1}>\beta_{3}$\}, whereas for
the time-dependent increase in alcoholic intake, peers will be more
influential \{$\beta_{4}<\beta_{6}$\}. The question of interest is:
\emph{Which of the theories best fits the data?}
\QEDE
\end{example}

The researchers' hypotheses are in fact
informative, as they are hypotheses in
which one explicitly defines direction or (relative) strength of
relationships based on prior information for usage in confirmatory
data analysis. Informative
hypotheses have a direct connection to
model translations of theory. For instance, the researcher from
Example 2 would be interested in the following two hypotheses that
have been arrived at by translating substantive theories via
constraints on model parameters:
\begin{eqnarray*}
&H_{1}:& \{\beta_{1}>\beta_{2}\}, \beta_{3},
\{\beta_{4}>\beta_{5}\}, \beta_{6}\\\mbox{versus} &H_{2}:&
\{\beta_{1}>\beta_{2}\}, \{\beta_{1}>\beta_{3}\},
\{\beta_{6}>\beta_{4}>\beta_{5}\}.
\end{eqnarray*}
The pertinent question is: Given $H_1$ and $H_2$, which of the two
hypotheses has more support from the data?

A researcher might bring the classical or frequentist statistical
viewpoint to bear on the central question of interest. One would
then normally proceed to specify the traditional null
hypothesis, which assumes that none of the
covariate variables are associated with the response variable of
interest against the alternative that
at least one covariate variable is associated with the response
 variable:
\begin{displaymath}
H_{0}: \mbox{all} ~\beta_{i}~ \mbox{equal} ~0 \,\,\,\,\,\mbox{versus}
\,\,\,\,\,H_{3}: \mbox{not all} ~\beta_{i} ~\mbox{equal} ~0.
\end{displaymath}
There are several problems related to this procedure that leads one
to infer little information regarding the actual hypotheses of
interest, being $H_1$ and $H_2$. Generally, in the usual frequentist
sharp null hypothesis test setting, the
researcher often starts from the idea that $H_{3}$ holds and then
tests $H_{0}$ using an appropriate test statistic. If we assume
$\vet{\beta}$, the vector containing all $\beta_{i}$, is
$\vet{\delta}$ away from the zero vector $\vet{0}$, with
$\vet{\delta}>\vet{0}$ but very small, then by the consistency of
the testing procedure, the rejection of $H_{0}$ becomes the sure
event for $N$ sufficiently large \cite{PRESS03}. One could then
actually choose $N$ in accordance with the rejection of $H_{0}$.
More specifically, if the null hypothesis is
rejected, no information is gained regarding the fit of the
inequality constrained hypothesis of interest. Note that the
research questions of actual interest are not directly incorporated
into the alternative hypothesis. Post
hoc directional tests are then usually employed with certain
corrections on the maintained significance level to assess the
inequalities deemed interesting in the actual research hypothesis.
If one considers $H_{1}$ above, these post hoc tests would amount to
assessing:
\begin{equation}\label{seq}
\begin{array}{ccc}
&H_{01}:& \beta_{1}=\beta_{2} \,\,\,\,\,\mbox{versus} \,\,\,\,\,H_{11}:
 \beta_{1}-\beta_{2}>0\\
\mbox{and} &H_{02}:& ~\beta_{4}=\beta_{5} \,\,\,\,\,\mbox{versus}
\,\,\,\,\,H_{12}: \beta_{4}-\beta_{5}>0.
\end{array}
\end{equation}
The researcher is left with the situation in which several test
results (those for the omnibus test and the post hoc tests) have to
be combined to evaluate a single model translated theory. Such a
situation may eventually force the researcher to make arbitrary
choices. For example, how would one evaluate the situation where not
all directional alternatives are accepted, or when the rather
arbitrary significance threshold is surpassed by an arbitrarily
small amount? Such problems abound especially in the social sciences
where it is not uncommon to find situations where power is
sufficient for obtaining significance somewhere while being
insufficient to identify any specific effect \cite{MAX04}. The power
gap between a single test and a collection of tests often renders
the situation in which the omnibus test proves significant in the
sense that the obtained \emph{p}-value is smaller than or equal to
the pre-specified significance level, while the individual post hoc
tests lack power such that successive testing efforts may find
erratic patterns of ``significant'' \emph{p}-values.

If the null hypothesis is not rejected when
testing $H_0$ against $H_3$, there is still a possibility that it
could be rejected when testing it against the hypotheses of
interest, namely $H_1$ and $H_2$. Inequality
constraints contain information, in
the form of truncations of the parameter space, and when properly
incorporated, more efficient inferences can result. To gain power,
one could therefore specify inequality constrained alternatives more
in tune with substantive theoretical beliefs, instead of the
traditional alternative $H_{3}$. This
way the null hypothesis, if rejected, will be
rejected in favor of the constrained alternative. Our researcher
would then embark on testing
\begin{eqnarray}\label{ineseq}\nonumber
&H_{0}:& \beta_{1}= \beta_{2}= \beta_{4}= \beta_{5}=0\\\mbox{versus}
&H_{4}:& \beta_{1}-\beta_{2}\geqslant0,~
\beta_{4}-\beta_{5}\geqslant0, ~and \nonumber\\&&\beta_{1},~
\beta_{2},~ \beta_{4}, ~\mbox{and} ~\beta_{5} ~\mbox{do not all
equal}~0
 \nonumber\\
\mbox{and}~~~~~~~~~~\\
&H_{0}:& \beta_{1}= \beta_{2}= \beta_{3} = \beta_{4}= \beta_{5}=
\beta_{6} = 0 \nonumber\\\mbox{versus} &H_{5}:&
\beta_{1}-\beta_{2}\geqslant0, ~\beta_{1}-\beta_{3}\geqslant0,
~\beta_{6}-\beta_{4}\geqslant0, ~\beta_{4}-\beta_{5}\geqslant0, ~and
\nonumber\\&&\beta_{1}, ~\beta_{2}, ~\beta_{3}, ~\beta_{4},
~\beta_{5} ~\mbox{and} ~\beta_{6} ~\mbox{do not all
equal}~0\nonumber
\end{eqnarray}
respectively, in order to convey more information regarding the
model translated theories of interest. Yet again, there are certain
problems that render the information to be inferred from these
omnibus tests to be limited.

First, there is an important difference between tests of the form
(\ref{ineseq}) and tests of the form (\ref{seq}). The former states
that a directional effect is present when the
alternative is accepted, but it does
not give which of the constituent directional effects is
significant. For such an evaluation one needs to resort to tests of
the latter form, which takes us back to the problems associated with
combining several test results to evaluate a single model translated
theory as discussed earlier. Moreover, for complex models and
multivariate settings there may not generally be optimal solutions
for frequentist inequality constrained testing alternatives such as
those in (\ref{ineseq}). The interested reader is referred to
\cite{BBBB72, SILSEN05} for overviews on the possibilities of
frequentist inequality constrained hypothesis testing. But even if
these frequentist alternatives were available, the researcher would
still run into a problem when wanting to evaluate which theory or
plausible model fits the data best. One possibility is to test the
null hypothesis against each of the theories
in the form of inequality constrained
alternatives. This would help one to
obtain some evidence for the support for each of the separate
theories, but it would still not answer the question concerning
which theory is best. It is very well possible that in all of the
tests the null hypothesis is rejected in
favor of the inequality constrained alternative.

To assess the researchers' substantive theory in light of the
available data, one needs to directly compare the constrained
alternatives. This involves the simultaneous evaluation of multiple
model translated theories, and for such an exercise, no frequentist
possibilities are available. Therefore, Bayesian model
selection is posed as an alternative to
hypothesis testing. Posterior probabilities can be computed for all models under consideration,
which enables the direct comparison of both nested and non-nested
models. The incorporation of inequality constrained theory
evaluation in a Bayesian computational framework has been formulated
for multilevel models in \cite{KHOI06}. In
the next section it will be shown how the inequality constrained
multilevel linear model can be given a Bayesian formulation, how the
model parameters can be estimated using a so-called augmented Gibbs
sampler, and how posterior probabilities can be
computed to assist the researcher in model selection. Those wishing
to skip this section may find general information regarding Bayesian
estimation and model selection in Chapters 3
and 4. Subsequently, the two examples described above will be
analyzed in the inequality constrained Bayesian framework to
elaborate model selection among competing
inequality constrained model translated theories. This will be done
in Sections \ref{ch11_sec4} and \ref{ch11_sec5}. The chapter will be
concluded with a discussion in Section \ref{ch11_sec6}.

\section{Bayesian Estimation and Model Selection}
\label{ch11_sec3}
\subsection{Introduction}
\label{ch11_sec3intro}

In Bayesian analysis, model specification has two parts to it:
\begin{enumerate}
\item The likelihood function $f(\vet{D}|\vet{\theta})$, which defines
 the probability distribution of the observed data $\vet{D}$
 conditional on the
unknown (model) parameters $\vet{\theta}$
\item The prior distribution $p(\vet{\theta})$ of the model parameters
 $\vet{\theta}$.
\end{enumerate}
Bayesian inference proceeds via
specification of a posterior distribution $p(\vet{\theta}|\vet{D})$ for $\vet{\theta}$, which is
obtained by multiplying the likelihood and the prior distribution:
\begin{equation}
p(\vet{\theta}|\vet{D}) =
\frac{f(\vet{D}|\vet{\theta})p(\vet{\theta)}}{m(\vet{D})} \propto
f(\vet{D}|\vet{\theta})p(\vet{\theta}),
\end{equation}
where $m(\vet{D})$ is the marginal distribution of $\vet{D}$. The
posterior distribution
$p(\vet{\theta}|\vet{D})$ contains the state of knowledge about the
model parameters given the observed data and the knowledge
formalized in the prior distribution. Random draws from the
posterior distribution are then used
for inferences and predictions. In the sequel it will be explained
how samples can be drawn from the posterior
distribution.

For (\ref{vectwolevelmodel}), the likelihood $f(\vet{D}\mid\vet{\theta})$ is
\begin{equation}
\prod_{j=1}^{J} \int_{\vet{u}_{j}} \left\{ \prod_{k=1}^{N_j}
\frac{1}{\sqrt{2\pi}\sigma}
\exp\left(-\frac{(y_{kj}\!-\!\vet{x}_{kj}
\vet{\beta}^T\!-\!\vet{z}_{kj}\vet{u}^{T}_{j})}{2\sigma^{2}}\right)
\right\} p(\vet{u}_{j}\!\mid\!\vet{0},\vet{V})~d\vet{u}_{j},
\label{likelihood}
\end{equation}
where $\vet{D}$ = ($\vet{y}_{kj},\vet{x}_{kj},\vet{z}_{kj}:
k=1,\dots,N_j;~j=1,\dots,J$), $\vet{\theta}$ = ($\vet{\beta}$,
$\vet{V}$, $\sigma^2$), and $p(\vet{u_{j}} \mid \vet{0},\vet{V})$ is
a normal distribution with
mean $\vet{0}$ and covariance matrix $\vet{V}$.

Suppose we have a total of $S$ competing hypotheses or model
translated theories $H_{s}$ for $s=1,\ldots,S$, where $H_1$ is the
encompassing model (in the remainder of the text we will use the
terms ``hypothesis'' and ``model'' interchangeably). The
encompassing model is one where no constraints are put on the
(model) parameters and therefore all other models are nested in
$H_1$. If $p(\vet{\theta}|H_{1})$ denotes the prior distribution of
$H_1$, then it follows that the prior distribution of $H_s$ for
$s=2,\dots,S$ is
\begin{equation}
p(\vet{\theta}|H_{s}) =
\frac{p(\vet{\theta}|H_{1})I_{\vet{\theta}\in{H_{s}}}} {\displaystyle\int
p(\vet{\theta}|H_{1})I_{\vet{\theta}\in{H_{s}}} d\vet{\theta}}.
\label{indicatoreqn}
\end{equation}
The indicator function $I_{\vet{\theta}\in{H_{s}}}=1$ if the
parameter values are in accordance with the restrictions imposed by
model $H_s$, and $0$ otherwise. Equation (\ref{indicatoreqn})
indicates that for each model under investigation, the constraints
imposed on the model parameters are accounted for in the prior
distribution of the respective model. Using independent prior
distributions for each of the model parameters, the prior
distribution of the unconstrained encompassing model $H_1$ can be written as the product
\begin{equation}
p(\vet{\theta}|H_{1}) = p(\vet{\beta})\times p(\vet{V})\times
p(\sigma^2), \label{encompassingprior}
\end{equation}
where $p(\vet{\beta})$, $p(\vet{V})$, and $p(\sigma^2)$ are the
prior distributions of $\vet{\beta}$, $\vet{V}$, and $\sigma^2$,
respectively. In order to obtain a conjugate model specification,
normal priors will be used
for the fixed effects $\vet{\beta}$, a scaled inverse
$\chi^{2}$
prior for $\sigma^2$, and an inverse Wishart prior for $\vet{V}$. It follows that
for the unconstrained encompassing model $H_1$, the posterior
distribution of the parameters in
$\vet{\theta}$ is proportional to the product of (\ref{likelihood})
and (\ref{encompassingprior}).

In what follows, it is explained how prior distributions for
$\vet{\beta}$, $\vet{V}$, and $\sigma^2$ will be specified. As
mentioned in Chapter 4 (see also \cite{KHOI06, KKHOI05}), the
encompassing prior should not
favor the unconstrained or any of the constrained models. Because
all constraints are on the parameters in the vector $\vet{\beta}$,
each of the $\beta$s will be assigned the same prior distribution.
In general, the estimate for the regression coefficient $\beta_{0}$
in a linear regression model with no covariates,
$y=\beta_{0}+\varepsilon$, where $y$ is the dependent variable and
$\varepsilon$ is an error term, is the mean of $y$ (i.e.,
$\hat{\beta}_{0}=E(y)$). Each of the parameters in $\vet{\beta}$
will therefore be assigned a normal distribution with mean equal to the mean of the response
variable (from the data) and a large variance chosen so that the
prior has minimal influence on the posterior
distribution of the parameter. The
prior distribution of $\sigma^2$ will also be data based --
$\sigma^2$ will be assigned a scaled inverse
$\chi^{2}$-distribution with $1$ degree of freedom and scale equal to the variance of
the response variable. Lastly, $\vet{V}$ will be assigned an inverse
Wishart prior
distribution with $R+1$ degrees of freedom and as scale matrix the
$R$$\times$$R$ identity matrix where $R$ is the dimension of
$\vet{V}$. Estimating the covariance matrix $\vet{V}$ is challenging
especially when $R>2$. This is because each of the correlations
(between the components of $\vet{u}$ in (\ref{randomterms})) has to
fall in the interval $[-1,1]$ and $\vet{V}$ must also be positive
definite. Setting the degrees of freedom to $R+1$ ensures that each
of the correlations has a uniform distribution on $[-1,1]$
(\cite{GEHIL07}). Although setting the degrees of
freedom to $R+1$
ensures that the resulting model is reasonable for the correlations,
it is quite constraining for the estimation of the variance terms in
$\vet{V}$. Therefore, when $R>2$, it is recommended to model
$\vet{V}$ using a scaled inverse Wishart distribution. The interested reader is
referred to \cite{GEHIL07} for more details on the implementation.

\subsection{Estimation}
\label{ch11_sec3Esti}

In this section it is explained how samples can be obtained from the
posterior distribution of $H_1$ and
how they can be used for inferences. With conjugate
prior specifications, in
(\ref{encompassingprior}), the full conditional distributions of
$\vet{V }$ and $\sigma^2$ are inverse Wishart and scaled inverse $\chi^{2}$
distributions, respectively, and the full conditional distribution of
each parameter in the vector of fixed effects $\vet{\beta}$ is a
normal distribution.

The Gibbs sampler (see, for example,
\cite{GSL92, KHOI06, SMROB93}), which is an iterative procedure, can
be used to sample from the conditional distribution of each model
parameter -- the set of unknown parameters is partitioned and then
each parameter (or group of parameters) is estimated conditional on
all the others. To sample from the posterior
distribution of the encompassing model
$H_1$ described in Section \ref{ch11_sec3intro}, first initial
values are assigned to each of
the model parameters. Next, Gibbs sampling
proceeds in four steps, namely:

\begin{itemize}
\item Sample $\vet{u}_j$ for $j=1\dots,J$ from
$\mathcal{N}(\vet{\Phi}_j,\vet{\Sigma}_j)$ where
\begin{equation}
\vet{\Phi}_j=\frac{\vet{\Sigma}_j}{\sigma^2}\sum_{k=1}^{N_j}\vet{z}^T_{kj}\left(y_{kj}-\vet{x}_{kj}
\vet{\beta}^T\right)   \nonumber
\end{equation}
and
\begin{equation}
\vet{\Sigma}_j=\left[\frac{\sum_{k=1}^{N_j}\vet{z}^T_{kj}\vet{z}_{kj}}{\sigma^2}
+ \vet{V}^{-1}\right]^{-1}.    \nonumber
\end{equation}

\item If the prior distribution $p(\sigma^2)$ of $\sigma^2$ is an
inverse chi-square distribution with degrees of
freedom $\gamma$
and scale $\omega^2$, then sample $\sigma^2$ from a scaled inverse
$\chi^{2}$-distribution with degrees of freedom $\gamma + \sum_{j=1}^{J} N_j$ and
scale
\begin{equation}
\gamma\omega^2 + \sum_{j=1}^{J}\sum_{k=1}^{N_j}\left(y_{kj} -
\vet{x}_{kj}\vet{\beta}^T - \vet{z}_{kj}\vet{u}^T_j\right)^2.   \nonumber
\end{equation}
\item If the prior distribution $p(\vet{V})$ of $\vet{V}$ is an inverse
 Wishart distribution with
degrees of freedom $\lambda$ and scale matrix $\vet{T}$, sample \vet{V} from
an inverse Wishart
distribution with degrees of freedom $\lambda + J$ and scale matrix
\begin{equation}
\sum_{j=1}^{J}\vet{u}_{j}\vet{u}^T_{j} + \vet{T}.   \nonumber
\end{equation}
\item Let $\vet{\beta}$ = $\{\beta_1,\ldots,\beta_p,\ldots,\beta_P\}$.
 If the prior distribution of
$\beta_p$ is a normal distribution with mean $\mu_{p}$ and variance
$\tau_{p}^{2}$, then sample $\beta_p$ from a normal
distribution with mean
\begin{equation}
\frac{\frac{\mu_{p}}{\tau_{p}^2}+\sigma^{-2}\sum_{j=1}^{J}\sum_{k=1}^{N_j}
\left[y_{kj}- \sum_{\substack{i=1 \\ i\neq p}}^{P}
\beta_{i}x_{ikj}-\sum_{q=1}^{Q}u_{qj}z_{qkj}
\right]x_{pkj}}{\tau_{p}^{-2}+
\sigma^{-2}\sum_{j=1}^{J}\sum_{k=1}^{N_j}x_{pkj}^{2}}\nonumber
\end{equation}
and variance
\begin{equation} \left[\frac{1} {\tau_{p}^2} +
\frac{\sum_{j=1}^{J} \sum_{k=1}^{N_j} x_{pkj}^{2}}
{\sigma^2}\right]^{-1}. \nonumber
\end{equation}
\end{itemize}
\begin{sloppypar}
Effectively, the Gibbs sampler starts with
initial values for all the
parameters and then updates the parameters in turn by sampling from
the conditional posterior distribution of each parameter. Iterating
the above four steps produces a sequence of simulations
$\vet{u}^{(1)}_1,\ldots,\vet{u}^{(1)}_J$, $\sigma^{2(1)}$,
$\vet{V}^{(1)}$, $\beta^{(1)}_1,\ldots,\beta^{(1)}_{P}$;
$\vet{u}^{(2)}_1,\ldots,\vet{u}^{(2)}_J$, $\sigma^{2(2)}$,
$\vet{V}^{(2)}$, $\beta^{(2)}_1,\ldots,\beta^{(2)}_{P}$;
$\vet{u}^{(3)}_1,\ldots,\vet{u}^{(3)}_J$, $\sigma^{2(3)}$,
$\vet{V}^{(3)}$, $\beta^{(3)}_1,\ldots,\beta^{(3)}_{P}$; and so on
until the sequence has converged. The first set of iterations,
referred to as the \emph{burn-in}, must
be discarded since they depend on the arbitrary starting values. See
Chapter 3 and references therein for more information on convergence
diagnostics for the Gibbs sampler.
\end{sloppypar}

After convergence, samples drawn
from the posterior distribution can be
used to obtain parameter estimates, posterior standard deviations,
and central credibility intervals. See,
for example, \cite{HOI00}. To elaborate, suppose that $\vet{\beta} =
(\beta_1, \beta_2)$ and we have a sample ${(\beta^{(b)}_{1},
\beta^{(b)}_{2}), b=1,\ldots,B}$, from the posterior
distribution. To estimate the
posterior mean of $\beta_{1}$, a researcher would use
\begin{equation}
\frac{1}{B} \sum_{b=1}^{B} \beta^{(b)}_{1},
\end{equation}
and a $95\%$ central credibility interval (CCI) for $\beta_{1}$ would be obtained by taking the
empirical $.025$ and $.975$ quantiles of the sample of
$\beta^{(b)}_{1}$ values. Furthermore, estimates of functions of
parameters can also be obtained. For instance, suppose an estimate
for the posterior mean of $\beta_1-\beta_2$ and a credibility
interval is required. This is easily obtained by taking the
difference $\beta^{(b)}_{1}-\beta^{(b)}_{2}$, $b=1,\ldots,B$, and
using the computed values to obtain the posterior mean and
credibility interval. Samples from the posterior
distribution can also be used to draw
histograms to display the distributions of parameters and functions
of parameters.

\subsection{Model Selection}
  \label{ch11_sec3modelsel}

If $p(H_{s})$ and $m(\vet{D}|H_{s})$ denote the prior probability
and marginal likelihood of model $H_s$, respectively, then the
posterior model probability (PMP)
of $H_s$ is
\begin{equation}
\mbox{PMP}(H_s~|~\vet{D}) =
\frac{m(\vet{D}~|~H_s)p(H_s)}{\sum_{s^{\prime}=1}^{S}m(\vet{D}~|~
H_{s^{\prime}})p(H_{s^{\prime}})}. \label{posteriorprobability1}
\end{equation}
The method of encompassing priors (see \cite{KHOI06,KKHOI05} and Chapter
4), can be used to obtain posterior probabilities for each model
under investigation. If $1/c_s$ and $1/d_s$ are the proportions of
the prior and posterior distributions
of $H_1$ that are in agreement with the constraints imposed by model
$H_s$, then the Bayes factor $BF_{s1}$ comparing
$H_s$ to $H_1$ is the quantity $c_s/d_s$. Note that for each
constrained model $H_s$, the quantities $1/c_s$ and $1/d_s$ provide
information about the \emph{complexity} (``size'' of the parameter
space) and \emph{fit} of $H_s$, respectively. Subsequently, if $H_1$
is the encompassing model and assuming that each model $H_s$ is a
priori equally likely, it follows that
\begin{equation}
\mbox{PMP}(H_s|\vet{D}) = \frac{BF_{s1}}
{BF_{11}+BF_{21}+\cdots+BF_{S1}}, \label{posteriorprobability2}
\end{equation}
for each $s=1,\ldots,S$ and $BF_{11}=1$. In practice, therefore, one
only needs to specify the prior distribution and correspondingly the
posterior distribution of the
encompassing model. Next, samples are drawn from the specified prior
and posterior distributions, which are
then used to determine the quantities $1/c_s$ and $1/d_s$.
Subsequently, posterior probabilities can be computed using
(\ref{posteriorprobability2}) and the model with the highest
posterior probability is considered to be the one that gets the
highest support from the data. If the model with the highest
posterior probability is one of the constrained models, then
parameter estimates for the model can be obtained using the Gibbs
sampling procedure presented in
Section \ref{ch11_sec3Esti} with an extra step, namely that the
$\beta$'s are sampled from truncated normal
distributions (see
Chapter 3).

Note that if a diffuse
encompassing prior is used,
then for the class of models with strict inequality
constraints, such as
$\beta_1>\beta_2>\beta_3$ or $\beta_4>0$, the PMPs obtained will not be sensitive to the prior
specification. However for models with equality constraints, such as
$\beta_1=\beta_2 =\beta_3$ or $\beta_4=0$, PMPs strongly depend on the actual specification of
the encompassing prior. For
details on this, the interested reader is referred to Chapter 4 and
\cite{KHOI06, KLHOI07, KKHOI05}. In this chapter, models with
equality constraints are not considered, so sensitivity of
PMPs to the choice of
encompassing prior is not an
issue.

\section{School Effects Data Example}
\label{ch11_sec4}
\subsection{Data}
\label{ch11_sec4dat}

\begin{sloppypar}
The data used in this section are a subsample of the 1982 High
School and Beyond Survey.\footnote{This data collection provides the
second wave of data in a longitudinal, multi-cohort study of
American youth conducted by the National Opinion Research Center on
behalf of the National Center for Education Statistics. In the first
wave, conducted in 1980, data were collected from 58,270 high school
students and 1015 secondary schools by self-enumerated
questionnaires, personal and telephone interviews, and mailback
questionnaires.} It includes information on 7,185 students nested
within 160 schools. Data were obtained from \url{http://www.ats.ucla.edu/stat/paper examples/singer/default.htm}.
\end{sloppypar}

The data set includes the following variables:
\begin{enumerate}
\item \vet{mathach}: The response variable, which is a standardized
measure of mathematics achievement. The
variable $mathach$ has mean $12.75$, standard deviation $6.88$, and
range $-2.83$ to 24.99.
\item \vet{ses}: A composite and centered indicator of student
socioeconomic status. It was a composite of parental education,
parental occupation, and income. The variable \emph{ses} has mean
0.00014, standard deviation $0.78$, and range $-3.76$ to 2.69.
\item \vet{minority}: A student level dummy variable that
 was coded as
 $1$ if the student belonged to a minority and $0$ otherwise. Numbers
 of
minority and nonminority students were 1974 and 5211, respectively.
\item \vet{meanses}:  School level variable indicating the
 average of
student \emph{ses} values within each school. As \emph{ses} was
centered around its mean a score of 0 can be interpreted as
indicating a school with average (in fact average average) student
\emph{ses} values, whereas $-1$ and 1 indicate schools with below
average and above average student \emph{ses} values respectively.
The variable \emph{mses} has mean $0.0061$, standard deviation
$0.41$, and range $-1.88$ to 0.83.
\item \vet{sector}: School level dichotomous variable
 where $1$
indicates a Catholic school and $0$ indicates a public school.
Numbers of Catholic and public schools were 70 and 90, respectively.
\end{enumerate}
Let $mathach_{kj}$ and $ses_{kj}$ respectively represent the math
achievement and student socioeconomic
status for the \emph{k}th $(k = 1,\ldots, 7185)$ student in the
$j$th school $(j = 1,\ldots,160)$. Let $min_{j}$ be an indicator
variable defined to be 1 if subject $k$ in school $j$ belongs to an
ethnic minority, and 0 otherwise. Furthermore, let $cat_{j}$ and
$pub_{j}$ be school level indicator variables defined to be 1 if a
school is Catholic or public, respectively, and $0$ otherwise. It
should be noted that the variable $cat$ is equivalent to the
original variable \emph{sector}. The reason for defining a new
indicator variable \emph{pub} is because in a regression model, this
will make it possible to estimate the regression coefficients
corresponding to the covariates $cat$ and $pub$ and their
interactions with other covariates rather than estimating contrasts.
Furthermore, defining variables in this way enables one to put
constraints on the model parameters. Finally, let $mses_{j}$
represent the continuous school level variable \emph{meanses}.

\subsection{Theory and Models}
\label{ch11_sec4theory}

\begin{sloppypar}
Research into child and adolescent mathematical
achievement has spurred a vast stream
of sociological, psychological, and educational literature; see, for
example, \cite{ BEK93, BRYR99, GAM92, GEAR94, SING98}. Van den Berg,
Van Eerde, and Klein \cite{BEK93} conducted research into the
mathematical skills of ethnic minorities in the Dutch elementary
school system. They concluded that children from ethnic minorities
have less mathematical ability/maturity than children from the
native Dutch population. These effects were, in their view,
attributable to a language barrier and the differential use of
educational skills between the home and the school environment.
These effects are expected to persist throughout high school.
Gamoran \cite{GAM92} found that Catholic schools produce higher
overall math achievement in comparison
to public schools. The (partial) explanation for this was found in
the manner in which Catholic schools implement academic tracking. In
addition, \cite{BRYR99, SING98} have indicated that higher math
achievement occurs in schools where the
average student socioeconomic status is higher. It is these
expectations we want to express in a set of informative
hypotheses.
\end{sloppypar}

Assuming a linear relationship between a student's mathematics
achievement, \emph{ses} and \emph{min},
the relationship can be modeled using
\begin{displaymath}
mathach_{kj} = \pi_{1j} + \pi_{2j}ses_{kj} + \pi_{3j}min_{kj} +
\varepsilon_{kj},
\end{displaymath}
where
\begin{eqnarray*}
\pi_{1j}&=&\beta_{1}cat_{j}+\beta_{2}pub_{j}+\beta_{3}mses_{j}+u_{1j},\\
\pi_{2j}&=&\beta_{4}cat_{j}+\beta_{5}pub_{j}+\beta_{6}mses_{j}+u_{2j},
 \\
\pi_{3j}&=& \beta_{7},
\end{eqnarray*}
and with
\begin{displaymath}
\vet{u}=(u_{1j}, u_{2j})^T \sim \mathcal{N}(\vet{0}, \vet{V}),
~~\varepsilon_{kj} \sim \mathcal{N}(0, \sigma^{2}).
\end{displaymath}
Thus, the school-specific intercepts ($\pi_{1j}$) and $ses$ effects
($\pi_{2j}$) are related to the type of school and average
socioeconomic status of the school. Note that the coefficient
$\pi_{3j}$ does not vary across schools. To keep things simple we
are assuming it has the same value $\beta_7$ for each school $(j =
1,\ldots,160)$. Making the coefficient differ for each school, say
by having $\pi_{3j}$ = $\beta_{7}+u_{3j}$, would give rise to a
$3\times3$ covariance matrix $\vet{V}$ for $\vet{u}=(u_{1j},
u_{2j},u_{3j})^T$. Effectively, the extra term $u_{3j}$ introduces
three new variance components, namely cov($u_{1j},u_{2j}$),
cov($u_{2j},u_{3j}$), and var($u_{3j}$) that have to be estimated
from the data.

The following competing inequality constrained model translated
theories will be compared:
\begin{eqnarray*}
H_{1}&:& \beta_{1}, ~\beta_{2}, ~\beta_{3}, ~\beta_{4}, ~\beta_{5},
~\beta_{6}, ~\beta_{7}, \\
H_{2}&:& \{\beta_{1}>\beta_{2}\}, ~\beta_{3}, ~\beta_{4},
~\beta_{5}, ~\beta_{6}, ~\beta_{7}, \\
H_{3}&:& \beta_{1}, ~\beta_{2}, ~\beta_{3}, ~\beta_{4}, ~\beta_{5},
~\beta_{6}, ~\beta_{7} < 0,\\
H_{4}&:& \{\beta_{1}>\beta_{2}\}, ~\beta_{3}, ~\beta_{4},
~\beta_{5}, ~\beta_{6}, ~\beta_{7} < 0,\\
H_{5}&:& \{\beta_{1}>\beta_{2}\}, ~\beta_{3},
~\{\beta_{4}<\beta_{5}\}, ~\beta_{6}, ~\beta_{7} < 0,  \\
H_{6}&:& \{\beta_{1}>\beta_{2}\}, ~\beta_{3},
~\{\beta_{4}>\beta_{5}\}, ~\beta_{6}, ~\beta_{7} < 0.
\end{eqnarray*}
Model 1 is the unconstrained encompassing model. Model 2 expresses
the idea that students in Catholic schools have higher math
achievement than those in public
schools $\{\beta_{1}>\beta_{2}\}$. Model 3 expresses the viewpoint
that students belonging to a minority will have lower math
achievement than students not belonging
to an ethnic minority. As $min_{j}$ is an indicator variable defined
to be 1 if subject $k$ in school $j$ belongs to an ethnic minority,
the previous expectation means that $\beta_{7}$ should be negative,
so that $\beta_{7}<0$. Model 4 combines the viewpoints in models 2
and 3, namely that student in Catholic schools perform better than
those in public schools and that students belonging to ethnic
minorities perform worse than those not belonging to an ethnic
minority. Model 5 expresses the viewpoints of model 4, with the
additional expectation that the slopes for \emph{ses} are higher in
public compared to Catholic schools $\{\beta_{4}<\beta_{5}\}$.
Lastly, model 6 expresses the viewpoints of model 4, with the
additional expectation that the slopes for \emph{ses} are higher in
Catholic compared to public schools $\{\beta_{4}>\beta_{5}\}$.

\subsection{Results}
As mentioned before, Bayesian analysis requires specification of
prior distributions for all unknown parameters in the encompassing
model ($H_1$). For all analyses diffuse priors were used. The regression coefficients
$\beta_1,\ldots,\beta_7$ were each given normal prior distributions with mean $12.75$ and
variance $10^4$ (that is, standard deviation $100$). What this means
is that each of the coefficients is expected to be in the range
($-87$, $113$), and if the estimates are in this range, the prior
distribution is providing very little information in the inference.
Because the outcome and all predictors have variation that is of the
order of magnitude $1$, we do not expect to obtain coefficients much
bigger than 20, so prior distributions with standard deviation $100$
are noninformative. The variance covariance matrix $\vet{V}$ was
given an inverse Wishart prior distribution with $3$ degrees of
freedom and as
scale matrix a $2\times2$ identity matrix. Lastly, $\sigma^2$ was
given a scaled inverse $\chi^{2}$ prior distribution with $1$
degree of freedom
and scale $47$.

To obtain posterior model probabilities for the competing models,
$200,000$ samples (after a burn-in of
$10,000$) were drawn from the prior and posterior
distributions of the encompassing
model ($H_1$), respectively. For each of the constrained models
$H_2,\ldots,H_6$, the proportion of samples from prior and posterior
in agreement with the constraints on $\vet{\beta}$ were used to
estimate the posterior probabilities of each model. Table
\ref{SchoolEffectsPostProbTable} shows the resulting estimated
posterior probabilities, which express prior knowledge (model
translated theories using inequality constraints) being brought up to date with empirical data. As can
be seen in Table \ref{SchoolEffectsPostProbTable}, $H_5$ gets most
support from the data suggesting that, on average, students in
Catholic schools have higher math achievement than those in public schools and that student level
socioeconomic status is positively associated with mathematics
achievement with public schools having
higher slopes than Catholic schools. This is in line with the
findings in \cite{SING98}. Lastly, model 5 also suggests that
students from an ethnic minority have lower math
achievement than those who are not from
a minority. These findings are similar to what was observed in a
sample of children from the Netherlands \cite{BEK93}. It is
worthwhile to note that models 2 and 3 are nested in model 5,
implying that in a sense there is more evidence to support model 5
than just the PMP of $0.47$.
Stated otherwise, if models $2$ and $3$ were not part of the
competing set of models, the PMP
of model $5$ would have been bigger than $0.47$.
\begin{table}[t]
\centering \caption{Posterior model probabilities}
\begin{tabular}{cc}
\hline\
Model & ~~~~PMP  \\
\hline
$H_1$ &~~~~0.059  \\
$H_2$ &~~~~0.117  \\
$H_3$ &~~~~0.118  \\
$H_4$ &~~~~0.235  \\
$H_5$ &~~~~0.471  \\
$H_6$ &~~~~0  \\
\hline
\end{tabular}
\label{SchoolEffectsPostProbTable}
\end{table}
Subsequently, estimates for parameters of model $H_5$ were obtained
using constrained Gibbs sampling. Posterior
distributions of the model parameters
were monitored for $20,000$ iterations after a burn-in of $10,000$ and were summarized by posterior means,
standard deviations, and $95\%$ central credibility
intervals. These are displayed in Table
{\ref{SchoolEffectsEstimatesTable}.
\begin{table}[t]
\centering \caption{Estimates for $H_5$}
\begin{tabular}{cccc}
\hline\
Parameter & ~~~Mean & ~~~\emph{SD} & ~~~$95\%$ CCI \\
\hline
 $\beta_{1}$  &~~~14.33   &~~~0.20   &~~~(13.93, 14.73)  \\
 $\beta_{2}$  &~~~12.67   &~~~0.19   &~~~(12.30, 13.03)  \\
 $\beta_{3}$  &~~~4.18   &~~~0.33   &~~~(3.53, 4.84)  \\
 $\beta_{4}$  &~~~1.16   &~~~0.18   &~~~(0.81, 1.51)  \\
 $\beta_{5}$  &~~~2.64   &~~~0.16   &~~~(2.32, 2.95)  \\
 $\beta_{6}$  &~~~0.98   &~~~0.30   &~~~(0.38, 1.57)  \\
 $\beta_{7}$  &~~~$-2.76$   &~~~0.19   &~~~($-3.14$, $-2.38$)  \\
 \hline
Var($u_{1j}$)         &~~~1.99   &~~~0.33   &~~~(1.42, 2.71) \\
Cov($u_{1j},u_{2j}$) &~~~$-0.04$   &~~~0.19   &~~~($-0.01$, 0.35) \\
Var($u_{2j}$)         &~~~0.24   &~~~0.12   &~~~(0.09, 0.54) \\
\hline
$\sigma^2$            &~~~35.88   &~~~0.61   &~~~(34.71, 37.09)  \\
\hline
\end{tabular}
\label{SchoolEffectsEstimatesTable}
\end{table}
Relating the estimates to the theories behind model $H_5$, it can be
concluded that controlling for all other predictors in the model:
\begin{enumerate}
\item Average predicted score for mathematics
 achievement
is higher for Catholic than public schools. The average predicted
mathematics achievement scores for
students who are not minorities in schools with $meanses=0$ are
14.33 and 12.67 for Catholic and public schools, respectively.
\item Students belonging to ethnic minorities have lower mathematics
achievement than those who are not from
minorities. The coefficient $\beta_7$ for $min$ implies that the
average predicted difference in mathematics
achievement scores between students
from minorities and nonminorities is 2.76.
\item Student level \emph{ses} is positively associated
with mathematics achievement with
public schools having higher slopes than Catholic schools; for
schools with average student $ses$ values (i.e., $mses=0$), each
extra unit of $ses$ corresponds to an increase of $2.64$ and $1.16$
in average mathematics achievement for
public and Catholic schools, respectively. Furthermore, in both
Catholic and public schools, the student level \emph{ses} effect on
math achievement increases with
increasing \emph{meanses}. Stated otherwise, the importance of $ses$
as a predictor for math achievement is
more pronounced for schools with higher values of $meanses$.

\end{enumerate}

\section{Individual Growth Data Example}
\label{ch11_sec5}
\subsection{Data}
\label{ch11_sec5dat}

As part of a larger study regarding substance abuse, Curran, Stice,
and Chassin \cite{CSC97} collected 3 waves of longitudinal data on
82 adolescents. Beginning at age 14, each year the adolescents
completed a 4-item instrument that sought to assess their
alcohol consumption during the
previous year. Using an 8-point scale (ranging from 0 = ``not at
all'', to 7 = ``every day''), the adolescents described the
frequency with which they (1) drank beer or wine, (2) drank hard
liquor, (3) had 5 or more drinks in a row, and (4) got drunk. The
data were obtained from \url{http://www.ats.ucla.edu/stat/examples/alda/}.

The dataset includes the following variables:
\begin{enumerate}
\item \vet{alcuse}: The dependent variable. This (continuous) variable
was generated by computing the square root of the mean of
participants' responses across its constituent variables (the
frequency with which the adolescents (1) drank beer or wine, (2)
drank hard liquor, (3) had 5 or more drinks in a row, and (4) got
drunk). The variable $alcuse$ has mean 0.92 and standard deviation
1.06 (range 0 to 3.61).
\item \vet{age}: Variable indicating age of adolescent.
\item \vet{peer}: A measure of alcohol use among the adolescent's
peers. This predictor was based on information gathered during the
initial wave of data collection. Participants used a 6-point scale
(ranging from 0 = ``none'', to 5 = ``all'') to estimate the
proportion of their friends who (1) drank alcohol occasionally and
(2) drank alcohol regularly. This continuous variable was generated
by computing the square root of the mean of participants' responses
across its constituent variables. The variable $peer$ has mean 1.02
and standard deviation $0.73$ (range 0 to 2.53)
\item \vet{coa}: A dichotomous variable where a $1$ indicates that an
adolescent is a child of an
alcoholic parent. Of the 246 adolescents, 111 are children of
alcoholic parents and the rest are children of nonalcoholic parents.
\end{enumerate}
Now let $alcuse_{kj}$ and $age_{kj}$ be the response (alcohol
use) and age, respectively, for
the \emph{j}th $(j = 1,..., 82)$ subject at age $k = 14, 15, 16$.
Next, let $t_{kj} = (age_{kj} - 14)/(2\times\mbox{std}(age))$, where
$\mbox{std}(age)$ denotes the standard deviation of $age$. It
follows that $t_{kj}=0$ corresponds to the baseline age of $14$.
Also, let $coa_{j}$ and $ncoa_{j}$ be indicator variables defined to
be 1 if the subject is the child of an alcoholic or not the child of
an alcoholic parent, respectively, and 0 otherwise. Additionally,
let $speer_{j}$ be the centered and scaled measure of alcohol use
among the adolescent's peers
obtained by subtracting the mean and dividing by two standard
deviations. In regression models that include both binary and
continuous predictors, scaling the continuous predictors by dividing
by $2$ standard deviations rather than $1$ standard deviation
ensures comparability in the coefficients of the binary and
continuous predictors \cite{GEL07, GEHIL07}. Note that for
interactions between two continuous variables, say $X_1$ and $X_2$,
each of the variables is scaled before taking their product; that
is, the interaction term is not obtained by scaling ($X_1\times
X_2$). It is the product of $(X_1-
\mbox{mean}(X_1))/(2\times\mbox{std}(X_1))$ and $(X_2-
\mbox{mean}(X_2))/(2\times\mbox{std}(X_2))$, where
$\mbox{mean}(X_r)$ and $\mbox{std}(X_r)$ denote the mean and
standard deviation of $X_r$, respectively.

\subsection{Theory and Models}              \label{ch11_sec5theory}

Previous longitudinal latent growth models have been used to examine
the relation between changes in adolescent alcohol
use and changes in peer alcohol
use. Curran, Stice, and Chassin \cite{CSC97} found that peer alcohol
use was predictive of increases in adolescent alcohol use.
Furthermore, Singer and Willett \cite{SWIL03} have shown that
adolescents with an alcoholic parent tended to drink more
alcohol as compared to those
whose parents were not alcoholics. Additionally, it is expected that
with regard to initial adolescent alcohol use, an alcoholic parent may be of more influence than
peers, whereas for rate of change with regard to alcohol intake,
peers may have more influence. It is these expectations we want to
investigate in a model and accompanying informative
hypotheses.

Assuming that the profiles of each subject can be represented by a
linear function of time, the model can be written as
\begin{displaymath}
alcuse_{kj} = \pi_{1j} + \pi_{2j}t_{kj} + \varepsilon_{kj},
\end{displaymath}
where
\begin{eqnarray*}
\pi_{1j}&=&\beta_{1}coa_{j}+\beta_{2}ncoa_{j}+\beta_{3}speer_{j}+u_{1j},\\
\pi_{2j}&=&\beta_{4}coa_{j}+\beta_{5}ncoa_{j}+\beta_{6}speer_{j}+u_{2j},
\end{eqnarray*}
and
\begin{displaymath}
\vet{u}=(u_{1j}, u_{2j})^{T} \sim \mathcal{N}(\vet{0}, \vet{V}),
~~\varepsilon_{kj} \sim \mathcal{N}(0, \sigma^{2}).
\end{displaymath}
Thus, the subject-specific intercepts ($\pi_{1j}$) and time effects
 ($\pi_{2j}$)
are related to peer alcohol use and whether parent(s) is/are alcoholic
 or not.

The following competing models will be compared:
\begin{eqnarray*}
H_{1}&:& \beta_{1}, ~\beta_{2}, ~\beta_{3}, ~\beta_{4}, ~\beta_{5},
~\beta_{6},\\
H_{2}&:& \{\beta_{1}>\beta_{2}\}, ~\beta_{3}, ~\beta_{4},
~\beta_{5}, ~\beta_6, \\
H_{3}&:& \{\beta_{1}>\beta_{3}\}, ~\beta_2,
~\{\beta_{4}<\beta_{6}\}, ~\beta_{5},  \\
H_{4}&:& \{\beta_{1}>\beta_{2}\}, ~\beta_{3},
~\{\beta_{4}>\beta_{5}\},  ~\beta_{6}.
\end{eqnarray*}
Model $1$ is the unconstrained model. Model $2$ expresses the theory
that adolescents with an alcoholic parent are more prone to higher
alcohol use at baseline
$\{\beta_{1}>\beta_{2}\}$. Model $3$ expresses the theory that with
regard to an adolescent's alcohol use, parents have more influence than peers at baseline
$\{\beta_{1}>\beta_{3}\}$, whereas over time peers have more
influence $\{\beta_{4}<\beta_{6}\}$. Model $4$ expresses the theory
that adolescents with an alcoholic parent are more prone to higher
alcohol use at baseline
$\{\beta_{1}>\beta_{2}\}$, as well as over time
$\{\beta_{4}>\beta_{5}\}$.

\subsection{Results}

The prior distributions for the parameters in the encompassing model
were specified as follows. The regression coefficients
$\beta_1,\ldots,\beta_6$ were each given normal prior distributions with mean $0.92$ and
variance $10^4$. The variance covariance matrix V was given an
inverse Wishart
prior distribution with $3$ degrees of freedom and a $2\times2$ identity matrix as
scale matrix. Turning to the prior on $\sigma^2$, we used a scaled
inverse $\chi^{2}$-distribution with $1$ degree of
freedom and scale
$1.12$. Subsequently, $200,000$ samples (after a burn-in of $10,000$) were drawn from the prior and the
posterior distributions of the
encompassing model, respectively. For each of the models $H_2$,
$H_3$, and $H_4$, the proportion of samples from prior and posterior
distribution of $H_1$ in agreement
with the constraints on $\vet{\beta}$ were used to estimate the
posterior probabilities of each model. These are displayed in Table
\ref{IndividualGrowthPostProbTable}.
\begin{table}[b]
\centering \caption{Posterior model probabilities}
\begin{tabular}{cc}
\hline\
Model & ~~~~PMP  \\
\hline
$H_1$ &~~~~0.208  \\
$H_2$ &~~~~0.416  \\
$H_3$ &~~~~0.000  \\
$H_4$ &~~~~0.375  \\
\hline
\end{tabular}
\label{IndividualGrowthPostProbTable}
\end{table}

The posterior probabilities suggest that the support in the data is
highest for model $H_2$. Subsequently, estimates for parameters of
model $H_2$ were obtained using constrained Gibbs
sampling. Posterior
distributions of the model parameters
were monitored for $20,000$ iterations after a burn-in of $10,000$ and were summarized by posterior means,
standard deviations, and $95\%$ central credibility
intervals, which are presented in Table
 {\ref{IndividualGrowthEstimatesTable}.
Looking at the PMPs for models
$2$ and $4$ in Table
 {\ref{IndividualGrowthPostProbTable} suggests
that model $4$ is not much worse than $2$. In Table
 \ref{IndividualGrowthEstimatesTable},
the estimate for $\beta_4$ is less than that of $\beta_5$; this is
opposite to the constraint $\beta_4>\beta_5$ of model $4$. This
suggests that the reason why model $2$ has a higher
PMP than model $4$ is because the
constraint on the parameters $\beta_4$ and $\beta_5$ in model $4$ is
not in accordance with the data, whereas model $2$ does not put any
constraints on these parameters.
\begin{table}[t]
\centering \caption{Estimates for $H_2$}
\begin{tabular}{cccc}
\hline\
Parameter & ~~~Mean & ~~~$SD$ & ~~~$95\%$ CCI \\
\hline
 $\beta_{1}$  &~~~0.97   &~~~0.11    &~~~(0.75, 1.19)  \\
 $\beta_{2}$  &~~~0.39   &~~~0.10   &~~~(0.19, 0.59)  \\
 $\beta_{3}$  &~~~1.01   &~~~0.15   &~~~(0.70, 1.31)  \\
 $\beta_{4}$  &~~~0.43   &~~~0.15   &~~~(0.15, 0.72)  \\
 $\beta_{5}$  &~~~0.45   &~~~0.13   &~~~(0.19, 0.72)  \\
 $\beta_{6}$  &~~~$-0.35$  &~~~0.20   &~~~($-0.74$, 0.04)  \\
 \hline
Var($u_{1j}$)         &~~~0.27   &~~~0.08   &~~~(0.14, 0.46)  \\
Cov($u_{1j},u_{2j}$)  &~~~$-0.01$   &~~~0.05   &~~~($-0.12$, 0.07) \\
Var($u_{2j}$)         &~~~0.18   &~~~0.05   &~~~(0.09, 0.29) \\
\hline
$\sigma^2$            &~~~0.35   &~~~0.05   &~~~(0.26, 0.46)  \\
\hline
\end{tabular}
\label{IndividualGrowthEstimatesTable}
\end{table}
Based on the estimates in Table
\ref{IndividualGrowthEstimatesTable}, the following can be
concluded:
\begin{enumerate}
\item Controlling for peer alcohol use, baseline (age = $14$),
adolescent alcohol use was higher
in children of alcoholics than in children with nonalcoholic
parents. The difference in average baseline alcohol use was
$\beta_1-\beta_2=0.58$ with $95\%$ central credibility
interval ($0.28, 0.88$).
\item Since $\beta_3$ is the coefficient for $speer=(peer -
 \mbox{mean}(peer))/(2\times\mbox{std}(peer))$, it follows
that the coefficient for the original variable $peer = 1.01 /
(2\times \mbox{std}(peer))= 0.69$. This implies that controlling for
whether or not a parent is alcoholic, for every point difference in
peer alcohol use, baseline adolescent alcohol
use is $0.69$ higher. Stated
otherwise, teenagers whose peers drink more at age 14 also drink
more at 14.
\item Adolescent alcohol use tended
 to increase over time at rates of
$\beta_4=0.43$ and $\beta_5=0.45$ per year for children of
alcoholics and nonalcoholics, respectively. However, there is no
difference between the rates, $\beta_4-\beta_5= -0.02$ with $95\%$
central credibility interval ($-0.41,
0.37$).
\item Since $\beta_6$ is the coefficient for the interaction between
$t_{jk}=(age-14)/(2\times \mbox{std}(age))$ and
$speer=(peer-\mbox{mean}(peer))/(2\times \mbox{std}(peer))$, it
follows that the coefficient for the interaction between $peer$ and
$age$ is $-0.35/(4\times\mbox{std}(peer)\times\mbox{std}(age)) =
-0.15$. However, the CCI for $\beta_6$ contains 0, so there is no
evidence to suggest that the coefficient is different from zero.
This implies that $peer$ alcohol use does not influence adoloscents'
alcohol use over time.
\end{enumerate}

\section{Discussion}    \label{ch11_sec6}
George Box is credited with the quote, ``all models are wrong, but
some are useful'' \cite{BOXDRA87}. A basic principle of scientific
inference is that a good fit of a model to a set of data never
proves the truth of the model. Indeed if one does find the best
fitting model, it may not be theoretically plausible or represent
the actual state of affairs. No (statistical) technique can prove
that a model is correct; at best, we can give evidence that a
certain model or set of models may or may not be a plausible
representation of the unobservable forces that generated the dataset
at hand.

There is, therefore, the possibility of the existence of unexplored
models that may yield superior posterior probabilities compared to
the set of models considered by a researcher. In practice, it would
be possible to evaluate all possible combinations of constraints in
the model set in order to obtain the best possible model given a
certain index of model fit. This however takes us into the
exploratory realm of data analysis,
which may tempt us into hypothesizing after results are known and,
as such, imposes physical as well as philosophical restrictions on a
meaningful scientific method.

The crux of a meaningful scientific method
is the exclusion of plausible alternatives. In the
exploratory mode many models are
included that may not be theoretically plausible or represent an
approximation of the actual state of affairs, even when they report
superior fit. Exploratory analysis
in our view, as a tool of scientific advance, predates the
scientific method in that it should be used
for developing ideas about relationships when there is little or no
previous knowledge. These ideas may then subsequently be tested in a
confirmatory analysis that adheres
to the scientific method.

The inequality constrained Bayesian approach to analysis of
multilevel linear models as advocated in this chapter explicitly
encourages researchers to formulate plausible competing theories for
confirmatory analysis and offers a
framework in which one is able to simultaneously evaluate all
possible alternative model translated theories with regard to model
fit and complexity. As such it has a strong connection with the
hypothetico-deductive scientific method and
the concept of strong inference \cite{PLAT64}. This method of
scientific advance has, coupled to inequality constrained Bayesian
confirmatory data analysis, the
following form (also see \cite{PLAT64}): (i) Devise on the basis of
previous knowledge (such as a former exploratory data
analysis on preliminary data,
previous results, or expert opinion) alternative theories. These
alternative theories will usually have inequality
constraints among the parameters of
its constituent hypotheses; (ii) devise a crucial experiment whose
possible outcomes will be able to demarcate maximally the
alternative theories or (when experiments are not possible)
establish which observational data one would need to exclude one or
more of the theories; (iii) perform the experiment or obtain the
observational data and establish the ``best model(s)'' with the
inequality constrained Bayesian confirmatory data
analysis framework; (iv) repeat
the cycle by refining the model(s) that remain(s) and/or by using
the outcome as prior knowledge in a natural process of Bayesian
updating.

In this chapter, we have considered only multilevel linear models.
However, the ideas presented in this chapter can be extended and
adapted to deal with multilevel logistic regression and other
multilevel generalized linear models. In such settings extra
complications are bound to arise because we are not dealing with
continuous data.

Furthermore, in situations in which the posterior probabilities are
similar or approximately equivalent for multiple models, the ``best
model'' question may not be most appropriate and one then may want
to embark on model averaging to take model uncertainty into account
in a stricter manner. Such issues may be the topic of further
research.

\section*{Acknowledgements}
This research was partly supported by grant NWO-VICI-453-05-002 of the Netherlands Organization for Scientific Research (NWO).
It was written while the first author was affiliated with the Twin Research and Genetic Epidemiology Unit of King's College London, UK, and while the second author was a Ph.D. candidate at the Department of Methodology and Statistics of Utrecht University, Utrecht, the Netherlands.
Parts of Section 6 are part of the second authors' unpublished Ph.D. thesis.

This version is a post-peer-review, pre-copyedit version of a chapter published as:
Kato, B.S., \& Peeters, C.F.W. (2008). Inequality Constrained Multilevel Models. In: H. Hoijtink, I. Klugkist, \& P.A. Boelen (Eds.). \emph{Bayesian Evaluation of Informative Hypotheses}.
New York: Springer, pp. 273--295.
The final authenticated version is available online at: \url{http://dx.doi.org/10.1007/978-0-387-09612-4_13}.

The authors would like to thank Judith Singer for indicating useful hierarchical datasets and the editors for useful comments that have improved this chapter.

\bibliographystyle{plainnat}

\begin{thebibliography}{20}
\bibitem[1]{BBBB72} Barlow, R.E., Bartholomew, D.J., Bremner, J.M.,
Brunk, H.D.: Statistical Inference under Order Restrictions: The
Theory and Application of Isotonic Regression. New York, Wiley
(1972)
\bibitem[2]{BEK93} Berg, W. van den, Eerde, H.A.A. van, Klein,
A.S.: Proef op de som: Praktijk en resultaten van
reken/wiskundeonderwijs aan allochtone leerlingen op de basisschool
[Practice and Results of Education Arithmatics and Mathematics for
Immigrant Children in Elementary School]. Rotterdam, RISBO (1993)
\bibitem[3]{BOXDRA87} Box, G.E.P., Draper, N.R.: Empirical
Model-Building and Response Surfaces. New York, Wiley (1987)
\bibitem[4]{BROWJ03}Browne, W.J.: MCMC Estimation in MLwiN (Version
2.0). London, Institute of Education University of London (2003)
\bibitem[5]{BRYR99} Bryk, A.S., Raudenbush, S.W.: Hierarchical Linear
Models: Applications and Data Analysis Methods. London, Sage (1999)
\bibitem[6]{CSC97} Curran, P.J., Stice, E., Chassin, L.: The relation
between adolescent and peer alcohol use: A longitudinal random
coefficients model. Journal of Consulting and Clinical Psychology,
\textbf{65}, 130--140 (1997)
\bibitem[7]{GAM92} Gamoran, A.: The variable effects of high
school tracking. American Sociological Review, \textbf{57}, 812--828
(1992)
\bibitem[8]{GEAR94} Geary, D.C.: Children's Mathematical Development:
Research and Practical Applications. Washington, DC, APA (1994)
\bibitem[9]{GSL92} Gelfand, A.E., Smith, A.F.M., Lee, T.M.: Bayesian
analysis of constrained parameter and truncated data problems using
gibbs sampling. Journal of the American Statistical Association,
\textbf {87}, 523--532 (1992)
\bibitem[10]{GEL07} Gelman, A.: Scaling regression inputs by dividing
by two standard deviations. Statistics in Medicine (in press)
\bibitem[11]{GEHIL07} Gelman, A., Hill, J.: Data Analysis Using
Regression and Multilevel/Hierarchical Models. Cambridge, Cambridge
University Press (2007)
\bibitem[12]{GOLD95} Goldstein, H.: Multilevel Statistical Models (2nd
edition). London, Edward Arnold (1995)
\bibitem[13]{HOI00} Hoijtink, H.: Posterior inference in the random
intercept model based on samples obtained with Markov chain Monte
Carlo methods. Computational Statistics, \textbf{15}, 315--336
(2000)
\bibitem[14]{HOX02} Hox, J.: Multilevel Analysis: Techniques and
Applications. London, Lawrence Erlbaum Associates (2002)
\bibitem[15]{KHOI06} Kato, B.S., Hoijtink, H.: A Bayesian approach to
inequality constrained linear mixed models: estimation and model
selection. Statistical Modelling, \textbf{6}, 231--249 (2006)
\bibitem[16]{KLHOI07} Klugkist, I., Hoijtink, H.: The Bayes factor for
inequality and about equality constrained models. Computational
Statistics \& Data Analysis, \textbf{51}, 6367--6379 (2007)
\bibitem[17]{KKHOI05} Klugkist, I., Kato, B., Hoijtink, H.: Bayesian
model selection using encompassing priors. Statistica Neerlandica,
\textbf{59}, 57--69 (2005)
\bibitem[18]{LONG93} Longford, N.T.: Random Coefficient Models.
London, Oxford University Press (1993)
\bibitem[19]{MAX04} Maxwell, S.E.: The persistence of underpowered
studies in psychological research: Causes, consequences, and
remedies. Psychological Methods, \textbf{9}, 147--163 (2004)
\bibitem[20]{PLAT64} Platt, J.R.: Strong inference. Science,
\textbf{146}, 347--353 (1964)
\bibitem[21]{PRESS03} Press, S.J.: Subjective and Objective Bayesian
Statistics: Principles, Models, and Applications (2nd edition). New
York, Wiley (2003)
\bibitem[22]{SILSEN05} Silvapulle, M.J., Sen, P.K.: Constrained
Statistical Inference: Inequality, Order and Shape Restrictions.
Hoboken NJ, Wiley (2005)
\bibitem[23]{SING98} Singer, J.D.: Using SAS PROC MIXED to fit
multilevel models, hierarchical models, and individual growth
models. Journal of Educational and behavioral Statistics,
\textbf{24}, 323--355 (1998)
\bibitem[24]{SWIL03} Singer, J.D., Willett, J.B.: Applied Longitudinal
Data Analysis: Modeling Change and Event Occurrence. New York,
Oxford University Press (2003)
\bibitem[25]{SMROB93} Smith, A.F.M., Roberts, G.O.: Bayesian
computation via the Gibbs sampler and related Markov Chain Monte
Carlo methods. Journal of the Royal Statistical Society, Series B,
\textbf{55}, 3--23 (1993)
\bibitem[26]{SNBOS99} Snijders, T., Bosker, R.: Multilevel Analysis:
An Introduction to the Basic and Advanced Multilevel Modeling.
London, Sage (1999)
\end{thebibliography}


\vspace{1cm}
\addresseshere

\end{document}